\documentclass[10pt,twocolumn,english,proof,endnotes]{article}
\usepackage{times}
\usepackage[latin1]{inputenc}
\usepackage{geometry}
\makeatletter
\usepackage{babel}
\makeatother
\begin{document}

\title{Enforcing Bulk Mail Classification}

\author{Evan P. Greenberg and David R. Cheriton\\
 Stanford University\\
 \{evang,cheriton\}@dsg.stanford.edu}

\maketitle
\begin{abstract}
Spam costs US corporations upwards of \$8.9 billion a year, and comprises
as much as 40\% of all email received \cite{spamstats}. Solutions
exist to reduce the amount of spam seen by end users, but cannot withstand
sophisticated attacks. Worse yet, many will occasionally misclassify
and silently drop legitimate email. Spammers take advantage of the
near-zero cost of sending email to flood the network, knowing that
success even a tiny fraction of the time means a profit. End users,
however, have proven unwilling to pay money to send email to friends
and family.

We show that it is feasible to extend the existing mail system to
reduce the amount of unwanted email, without misclassifying email,
and without charging well-behaved users. We require that bulk email
senders accurately classify each email message they send as an advertisement
with an area of interest or else be charged a small negative incentive
per message delivered. Recipients are able to filter out email outside
their scope of interest, while senders are able to focus their sendings
to the appropriate audience.
\end{abstract}

\section{Introduction}

Unsolicited email has become a real problem. Companies around the
globe are losing billions of dollars each year in the form of lost
productivity. And the spammers are profiting; 8\% of Internet users
bought something from spam in 2003 \cite{spamstats}.

Unfortunately, the numbers in this game favor the spammers. A marketer
can send billions of emails at negligible cost; to realize a profit,
only a handful need result in a sale. Every email which makes it through
our arsenal of spam filters carries with it a very real cost to the
recipient, who must spend time to identify and discard the message.
Spammers have strong incentives to evade or subvert any mechanism
we deploy, while the recipients bear the cost of dealing with any
email that leaks through.

Antispam solutions generally attempt either to reduce the leakage
by improving the quality of filtration or to reduce the profit potential
by imposing a cost for sending spam. Approaches which attempt the
former are in the unenviable position of not being able to safely
err on either side in their classification, as they risk either being
ineffectual or dropping desired email. In this paper, we will primarily
take the latter approach, of introducing penalties for sending spam,
and we will show that in our system:

\begin{itemize}
\item Legitimate email is never misclassified or dropped.
\item Well-behaved users do not pay to send email.
\item Legitimate mass mailing is supported.
\end{itemize}
Clearly it is also necessary to provide attack resistance, assuming
sophisticated spammers who will optimize their strategies for any
solution we deploy.

This is accomplished by requiring bulk email senders to accurately
classify each email message they send as an advertisement and its
area of interest or else be charged a small negative incentive per
message delivered to a recipient, a micro-penalty. This micro-penalty
multiplied by the scale required for spam makes it uneconomic to lie.
Reliable classification makes it feasible for email recipients to
filter out advertising emails that are outside their scope of interest.
We show that it is feasible to log all email so that violations can
be acted on quickly and securely while resisting attacks, supporting
opt-in mailing lists as well as direct person-to-person email, and
requiring only modest changes to existing email servers while being
economic to deploy to Internet scale.

\section{Tagging Bulk Mail}

We require senders to annotate all bulk email with a new header field,
\textsc{X-Bulk-Mail.} At this time we define two types of bulk mail:
advertising and mailing lists (described further in Section \ref{sec:Opt-in-Bulk-Mail}).
Advertising emails are tagged with an ADV leader followed by a comma-separated
list of interest groups. Interest groups are defined hierarchically,
with terms separated by dots, and can be defined organically as the
system evolves. For instance, an advertisement for suntan lotion might
include the header

\begin{description}
\item [\textsc{X-Bulk-Mail:}]\textsc{ADV: rec.sports.swimming,}\\
\textsc{rec.sports.sailing}
\end{description}
A user can easily establish filters to discard mail for categories
he isn't interested in. Ideally this would be provided as a service
by his ISP.

When email received is considered to have been misclassified, whether
because it is lacking the appropriate header or because it has been
designated with an inappropriate interest group, the user will forward
the email to his ISP using his email client's bounce or redirect command.
This preserves the full headers of the email, allowing the penalty
phase of the system to take effect.

In our system, we define personal email to be any email which does
not carry an \textsc{X-Bulk-Mail} tag. It is treated in the same way
as bulk mail, and is presumed to be of interest to all recipients.
Email which is not of interest, such as advertising, can be objected
to on the basis of misclassification.

\section{Penalizing Misbehavior}

Each email server is extended to log relevant header information from
the mail it forwards over a window of two weeks. Specifically, we
store the (cryptographically secure) hash of the \textsc{Date, To,
From,} and \textsc{Received} fields.

When an ISP receives a spam complaint, it can perform a lookup to
validate that the complaint references a genuine email. It then forwards
the complaint on to the previous relay in the path. It also requires
the upstream relay pay it the required micro-penalty. That relay will
recover the cost when it passes the complaint on.

The result of this process is as if the micro-penalty were levied
against the ISP of the sender by the ISP of the recipient. The sender's
ISP can choose how to deal with the complaint, but should at a minimum
include forwarding the complaint to the user and imposing some form
of sanction.

For this system of penalties to work, we must also establish a few
procedural elements. First, as stated before, we have a history of
forwarded email to work from to prevent fraudulent use of the penalty
system. We also refuse to forward email that is more than a week old.
An enforceable penalty process must be set up between SMTP peers,
and relays must refuse unknown connections. In section \ref{sec:Incremental-Deployment}
we discuss how an ISP might redirect unknown relays to more permissive
ingress points. Finally, ISPs must specify some form of ingress rate
limit, to bound their potential outstanding liability. The monthly
service fee paid by end users could be considered a bond against which
the user borrows to send email.

The impact of these constraints on typical clients is negligible.
Consider an ISP which charges \$30/mo for Internet access, and a micro-penalty
fee of \$0.10. The ISP limits customers to sending 100 emails per
week, and terminates accounts on receipt of 10 spam complaints over
a 3-month period. In this scenario, the ISP has a maximum outstanding
liability of \$20.90 per client, which is safely less than the monthly
service fee. The ISP could also choose to allow clients to reset their
spam counter in exchange for a \$1 fee. Clients would need to send
more than 100 emails in a one week or send 10 offensive emails before
noticing the constraints of the system. Upgraded accounts with an
explicit surety could be provided to address these issues for the
handful of customers who require greater flexibility.

To the spammer, however, even a \$0.10 fee is profound. As a reference
point consider DoubleClick, which provides legitimate marketers with
the tools needed to maintain high-quality lists of interested users.
For these lists in Q1 of 2004 \cite{DoubleClick}, DoubleClick customers
saw a revenue of \$0.23 per email. Few enterprises could afford to
spend 40\% of their revenue on marketing, and few spammers will have
returns anywhere close to that seen by the carefully culled lists
maintained by professional marketing companies.

\section{Attack Resistance}

For clarity of presentation, we have so far ignored the possibility
of malicious entities. It is critical, however, to demonstrate that
attackers cannot evade the system to send email without being held
accountable, and that they cannot subvert the system to charge innocent
users.

\subsection{Assigning Responsibility}

Suppose Alice receives spam, for which she forwards a complaint to
her ISP. There are only three possible attackers: the sender, some
host on the path, and the receiver (her ISP). There is no way to send
email in our record route environment without being on-path (unless
there is an on-path accomplice, which is itself an attacker).

Assuming that there are no misbehaving hosts on the path between Alice
and the sender, it is clear the spammer will be charged. Using a fake
return address will not help, since the complaint will use the unforgeable
portion of the \textsc{Received} header path to reach his ISP, which
has a clear economic incentive to correctly identify the true sender
as well as the means to do so, since it controls where and how email
enters its network. The complaint cannot be forwarded by the sender
to another host because it will not be present in the mail history
for that node.

Suppose there is some host on the path between Alice and the sender
which is misbehaving. It cannot refuse to validate the spam complaint
without jeopardizing its relationship with its downstream peer. The
complaint will, in turn, only be accepted by the upstream ISP towards
the true sender. Spam forged by this host would result in the host
receiving complaints which could not be forwarded to any {}``upstream''
peers.

This misbehaving host could hijack legitimate mail traffic by replacing
the body of the email. This would be detected by the endpoints of
the communication channel, and eventually the culprit uncloaked. This
more pernicious attack also requires the malignant host be on the
path between Alice and some other user she wants to talk with, which
is already unlikely given the short length of typical SMTP relay paths
and the relative trustworthiness of genuinely on-path hosts.

Finally, Alice's ISP could itself send her spam, or refuse to take
action against spammers within the local network. The free market
suggests that such ISPs will not last very long, as users who find
this behavior distasteful will simply take their business elsewhere.

\subsection{Protecting Innocent Users}

We also prevent malicious nodes from abusing the spam reporting system
itself. In this case, the attacker could be an endpoint, some ISP
on the path between two nodes, or an off-path entity.

A client can only complain about an email she actually receives because
ISPs validate all complaints against their histories. This also automatically
excludes reverse path forgery and makes it easy to prevent recipients
from filing more than one complaint per spam.

A malicious on-path node is prevented from forging complaints for
the same reasons as for an endpoint. It would be possible for the
node to generate spam complaints about email which it was only supposed
to forward. However, because the sender is informed when complaints
are lodged, this type of misbehavior can be easily detected.

A true off-path attacker would need to somehow derive the headers
of emails traversing the network in order to file spam complaints.
Guessing these would not be feasible, but the attacker might try to
snoop email sessions or have an on-path accomplice forward the relevant
data. However, it still would not be possible for the attacker to
introduce a spam complaint into the network at any of the well-behaved
relays, as these nodes would know the attacker was not a suitable
next hop for the original email.

\subsection{Other Issues}

It is conceivable that users might accidentally file spam complaints
against desirable email. Anecdotal evidence from the SPAM-L mailing
list \cite{spam-l} indicates that some of the spam reports generated
in AOL's Feedback Loop \cite{AOLFeedback} are a result of just this
effect. In general, lower rates of spam are likely to lead to less
accidental complaints. Simple interface improvements, such as requesting
confirmation or password entry may also help to further reduce this
effect. Finally, our future work on conversational indemnity should
also help reduce the impact of accidental reports.

The issue of zombie hosts is also of concern, because it violates
the assumption that email sent by a host is intentional by the user.
However, the amount of damage a single host can do is severely constrained
due to the requisite ingress rate limiting. Furthermore, the threat
of losing email privileges gives users a real incentive to notice
and promptly clean up infections. Finally, consumer ISPs could provide
as a service to customers some level of early notification and/or
prevention through signature analysis or other monitoring technique.
In the end, we must hold a user accountable for the behavior of his
computer.

\section{\label{sec:Opt-in-Bulk-Mail}Opt-in Bulk Mail}

Mailing lists are special cases of bulk mailers, because although
they may need to send a number of messages each day to thousands of
subscribers, they have explicitly been allowed to do so by the process
of subscription. In light of this, it does not seem practical for
the mailing list to assume liability for every subscriber. Nor does
it seem fair to charge the original sender for every ultimate recipient,
when the sender cannot necessarily determine in advance who those
individuals are. However, it is also clear that mailing lists are
a widely used, and therefore desirable, feature of the current email
system.

To solve this we introduce recipient-side whitelisting. When an opt-in
bulk sender (such as a mailing list) sends email, it flags it with
a special \textsc{X-Bulk-Mail} header, which consists of a LIST leader
followed by an identifier for the list. The identifier could either
be derived from the list name or could be a random nonce. For instance,

\begin{description}
\item [\textsc{X-Bulk-Mail:}]\textsc{LIST: freefood.348290}
\end{description}
SMTP relays along the path do not include entries in their histories
for such email. The final relay, however, will drop any email marked
as list mail which is not explicitly whitelisted by that recipient.
Whitelisting is done based on the list identifier and, optionally,
the reverse path to the mailing list's remailer.

This mechanism works in exactly the way we expect mailing lists to
operate. The act of subscription adds the user's email address to
the list of addresses on the list and also adds the list to the local
whitelist. Unsubscription is simply removing the mailing list from
the local whitelist. Additionally, we gain the ability to force withdrawal
from a list by removing the local whitelist entry.

From the viewpoint of the list operator, opt-in lists in our system
carry the benefit of guaranteed protection from liability, because
users cede their ability to complain by virtue of subscribing to the
list. List operators could abuse this power by filling the list with
spam, but the ability to force unsubscription means that users only
pay temporarily for their mistakes.

\section{Implementation}

As a reference implementation we have produced an SMTP proxy intended
to be placed in front of a site's existing infrastructure. From this
vantage point it is straightforward to inspect all messages entering
the network from an SMTP peer and to log the requisite hash. Similarly,
messages from internal users destined for the outside world can be
logged and rate controlled.

This is also a suitable location to deal with spam complaints. When
a user receives an undesirable message, he simply redirects the message
(this preserves the original headers) to \textsc{spamsink@localdomain,}
and the proxy intercepts this message. It strips off the downstream
headers, validates the message against its logs, and then forwards
the complaint upstream. In particular, note that it is not necessary
to log at every SMTP host; within a single domain of trust, only one
log point is necessary.

To avoid duplication of configuration information, our proxy is tightly
coupled to a slave SMTP server. Commands are passed directly through
to the downstream server, except where needed to preserve the semantics
of the SMTP protocol. The proxy can then determine whether a command
was successful by inspecting the server's response, and takes that
into account when updating its state. The upshot of this is that the
proxy server can depend upon the real server to enforce any policies
the site may have in place (for instance, the set of addressable destinations)
without needing any separate logic or configuration. The only notable
exception to this is user authentication, which the proxy needs to
know about in order to perform rate limiting. But it only makes sense
that the proxy should provide this, as existing servers lack the functionality.

The exact performance characteristics of the proxy will, naturally,
depend upon the load characteristics of the site it serves. We assume
an ISP such as Princeton University \cite{PtonStats} (which publishes
its SMTP traffic statistics online). A generous interpretation of
the statistics available shows that in the month of February, 2005,
Princeton saw at most a few thousand SMTP messages per minute. This
figure is roughly in line with CAIDA measurements of 1.6 million SMTP
sessions per day \cite{CAIDA}.

If we assume a constant rate of 3000 messages per minute, or 60 million
messages over the two-week window, our solution requires about 3 gigabytes
of storage for logging. Our experiments show that storing the log
entry for each email costs approximately 3.1msec on a single-proc
P4 3.2GHz with 10k RPM SATA drive; for the suggested load this means
the SMTP relay (or a dedicated logging machine) will spend 15\% of
its time dealing with hashes. Checking a spam complaint against the
database of log entries costs just 0.77 msec. Computing the hash for
an email costs just 39 usec.

Full performance results, including analysis of performance under
load bursts, are pending. However, the results shown here do not include
any attempts to optimize the logging mechanism; each write results
in synchronous disk I/O. Note that it should be possible to maintain
the entire 3GB of history in memory for reasonable cost, with periodic
writes to disk just to prevent data loss in the event of a crash.
Writing the whole hash table to disk takes less than 5 minutes, and
could of course be done in stages.

\section{\label{sec:Incremental-Deployment}Incremental Deployment}

Although we present this work in a context which assumes global deployment,
it is also incrementally deployable, by which we mean that there is
a benefit realized by ISPs who deploy this system, even if all of
their neighbors do not. Specifically, an ISP who deploys our proposal
can guarantee that it will never drop email for its customers, while
also moderating the risk that this will cause an increase in spam
load.

Assume some ISP chooses to deploy our system. It will then have two
categories of neighbors: those with whom it can negotiate contracts,
and those with whom it cannot. When an ISP can set up the contractual
arrangement we specified earlier, it need not care if the peer ISP
has also deployed the system or not, because that ISP has already
agreed to be accountable for all the email it sends.

When an ISP cannot set up a contractual relationship with another
ISP, it must instead treat that neighbor as a potential spammer. Specifically,
it should keep track of the amount of spam received from that specific
ISP so that exceeding some threshold (say 100 objectionable emails
in a month) will cause the peer to be cut off. Future email from that
peer can then be bounced, with an indication that the ISP has been
identified as a source of spam and a few remedies suggested (such
as switching to a compliant ISP or asking your ISP to become compliant).
Our deploying ISP does assume financial responsibility for any email
accepted from this peer ISP, particularly if it is not destined for
one of its direct customers. It is important to realize that there
is no requirement that one must accept email from anyone on the Internet;
as such these peering relationships can be pre-screened to eliminate
known spammers.

Additionally, during this transitional phase, ISPs can continue to
use traditional antispam techniques such as content-based filtration
or blacklisting. These techniques are not particularly useful in a
world where everyone has deployed our system, but they can be applied
against these {}``unaccountable'' peer ISPs. However, it should
be understood that to avoid silently dropping email, an indicative
error response must be sent back if the traditional methods cause
the email to be classified as spam.

In this paper we treat the SMTP reverse path is if there could be
any number of ISPs on the path between sender and recipient. In practice,
of course, the sender ISP will contact the recipient's ISP directly
(by performing a lookup of an MX record for the destination domain).
But it is impractical to assume that every pair of ISPs will be able
to negotiate (and enforce) the contractual relationship we require.
Rather, deployment of our system will result in baby ISPs signing
contracts with larger for the ability to participate in the global
exchange of email. Conveniently, there is enough flexibility in the
system of MX records to allow automatic traversal of the tree (or
several trees). When an unknown peer attempts to send email, the recipient's
server will simply terminate the session, at which point the sender
will happily try the next entry in the list. In this manner, email
ingress gradually moves from less permissive to more permissive, eventually
settling on the appropriate entry point.

\section{Related Work}

A number of different solutions to the spam problem have been suggested;
some have even seen widespread deployment. Unfortunately, the existing
literature has not accurately gauged the sophistication of the attackers,
leading to solutions which are not effective enough to stem the ongoing
flood of spam.

Content-based filtering solutions, particularly those based on machine
learning, have been the most effective anti-spam treatment to date.
But they will never be able to guarantee that they will not decide
to randomly drop legitimate email. This is an anathema in a world
where we are trying to push to five-nine reliability and beyond. Worse
yet, because these solutions will always leak some amount of traffic,
they fail to address the core issue that spamming will continue to
be profitable so long as a trickle of bandwidth exists. In general,
not only is content-based filtering dangerous because it leads to
false positives, it continually becomes less effective as spammers
learn how to construct emails to get around the filtering algorithms.

More than a decade ago, Dwork and Naor introduced the idea of using
computational stamps to price email \cite{PennyBlack}. This idea
has been respun in a variety of different clothes, including a recent
paper by Balakrishnan and Karger \cite{SpamIAm}. Unfortunately, such
solutions are unlikely to ever succeed, because users are simply not
willing to pay for a service they view as being free. And if the fee
is purely computational (as suggested by the original work), then
it is too easy to farm the computation off to unknowing users through
the use of a botnet, cooperative spyware, or even a background javascript
task on a popular webpage. Advances in reverse Turing tests such as
CAPTCHAs \cite{Captcha}, though appealing in concept as a method
of postage stamping will suffer from the same popularity problems.
Additionally, spammers have apparently found ways to offload this
processing by using the puzzle images as conditions for admission
to porn sites \cite{NoCaptchas}.

There is also a continual effort from within the Internet community
to maintain various blacklists, which are then used to drop spam.
In theory these should never produce false positives, as only mail
validated as spam is added to the list. But they cannot prevent letting
some spam through, and thus are not a true deterrent. Additionally,
lists driven by Internet telescopy will only work so long as spammers
are not aware of the set of trigger addresses. Lists driven by voluntary
reporting are even scarier; eventually spammers will find a way to
pollute these lists with false reports, at which point claims of never
dropping desirable mail go out the window. Explicit whitelist solutions
such as ChoiceMail \cite{ChoiceMail} provide an interesting twist.
They do provide the additional benefit of requiring email to have
a legitimate return address. However, the whitelist request mechanism
can itself be leveraged to spam. Reverse Turing tests, though better
suited to this application than email stamping, still suffer the same
issues of being sharable.

The sheer scale of the spam problem on the Internet today, coupled
with the technological savvy exhibited by spammers, leaves us with
little choice but to introduce accountability into the network, and
then establishing a system of economic disincentives for abusers.
Our solution does this by introducing logs within the network. The
Bonded Sender \cite{BondedSender} and SHRED \cite{SHRED} systems
suggest achieving this goal through the use of what are effectively
prepurchased cryptographic stamps belonging to some trusted third
party. The Bonded Sender solution, though, because it only tries to
target (legitimate) bulk mailers, is only able to solve part of the
problem. The SHRED proposal, recent research by Krishnamurthy and
Blackmond, does seem more promising. It is our belief that by providing
validation from within the network, we are able to achieve better
attack resistance. In particular, we demonstrate in this paper that
the economic disincentives of our system will always work against
spammers; solutions requiring trusted parties are likely to find it
difficult to provide the same level of assurance. Additionally, by
providing spammers with the ability to achieve legitimacy by classifying
their email, we simultaneously reduce the load on the penalty system
and give users better automated visibility into the email they receive.

\section{Future Work}

One way in which this system could readily be extended would be to
provide conversational indemnity. That is, include the ability to
negate a spam complaint by presenting evidence (in the form of a prior
email) that the other party initiated communication. This could be
validated by the ISPs at each end of the communication chain using
the same hash history table which we use to validate spam complaints.

This sort of conversation-based indemnity could also be used to reduce
the effect of accidental complaints. Additionally, this mechanism
could be used to make it safe to set up an email autoresponder (along
the lines of the unix \emph{vacation} program) without turning your
computer into an attack reflector. It might also be possible for ISPs
to use this mechanism to relax ingress limits on users, since a response
to an email would constitute an acknowledgment that the original email
was not spam. 

The challenge here is in ensuring that an attacker cannot leverage
the indemnity mechanism to evade penalty (for instance, by playing
with the 2-week history window or reusing the same refutation multiple
times). Additionally, the ISP should be able to automatically refute
a complaint on behalf of the user, but we do not want the ISP to maintain
copies of all of a user's email.

\section{Conclusion}

We describe a system in which bulk email senders are required to accurately
classify each email message they send or be charged a small negative
incentive for each recipient. A history of forwarded email receipts
maintained by relays along the path provides automated verification
for spam complaints by recipients, which are forwarded towards senders
to provide a robust, attack-resistant penalty mechanism.

We show that this system allows person-to-person and opt-in bulk email
for free, without ever misclassifying or dropping email, needs only
modest changes to the existing SMTP infrastructure, and remains scalable
to Internet scale. Novel contributions include the use of service
fees as sureties for typical consumers, the separation of and improved
protection for inherently opt-in mailing list bulk mail versus opt-out
lists, and the enforceability of voluntary classification by advertisers.

\end{document}